\documentclass[mathleft]{an}
\usepackage{graphicx}
\usepackage{times}
\begin{document}

\Pagespan{666}{}
\Yearpublication{2010}%
\Yearsubmission{2010}%
\Month{00}%
\Volume{000}%
\Issue{00}%

\title{Towards pulsation mode identification in 3-D:
theoretical simulations of line profile variations in roAp stars}

\author{O. Kochukhov\inst{1}\fnmsep\thanks{
  Corresponding author: \email{oleg.kochukhov@fysast.uu.se}\newline}
\and
E. Khomenko\inst{2,3}
}
\titlerunning{line profile variations in roAp stars}
\authorrunning{O. Kochukhov \& E. Khomenko}
\institute{
Department of Physics and Astronomy, Uppsala University, Uppsala 75120, Sweden
\and
Instituto de Astrof\'{\i}sica de Canarias, 38205,
C/ V\'{\i}a L{\'a}ctea, s/n, Tenerife, Spain
\and
Main Astronomical Observatory, NAS, 03680, Kyiv, Ukraine
}

\received{01 Apr 2010}
\accepted{01 Apr 2010}
\publonline{later}


\abstract{%
Time-resolved spectroscopic observations of rapidly oscillating Ap (roAp) stars
show a complex picture of propagating magneto-acoustic pulsation waves, with
amplitude and phase strongly changing as a function of atmospheric height. We
have recently conducted numerical, non-linear MHD simulations to get an insight
into the complex atmospheric dynamics of magnetic pulsators. Here we use the
resulting time-dependent atmospheric structure and velocity field to predict
line profile variations for roAp stars. These calculations use realistic
atmospheric structure, account for vertical chemical stratification and treat
the line formation in pulsating stellar atmosphere without relying on the
simplistic single-layer approximation universally adopted for non-radial
pulsators. The new theoretical calculations provide an essential tool
for interpreting the puzzling complexity of the spectroscopic pulsations in roAp
stars.
}

\maketitle

\section{Introduction}

Rapidly oscillating Ap (roAp) stars is a group of cool magnetic Ap stars pulsating
in high-overtone, non-radial modes with periods around 10~min. Excitation of these
pulsations and the physics of their propagation in the stellar envelopes and atmospheres
is closely connected to the presence of global magnetic fields of several kG strength
(e.g., Balmforth et al. 2001; Saio 2005).

Recent time-resolved spectroscopic observations of roAp stars (Kochukhov \& Ryabchikova 2001;
Mkrtichian, Hatzes \& Kanaan 2003; Ryabchikova et al. 2007) showed a remarkably complex and
diverse pulsational variability of spectral lines of different chemical elements. In
particular, one often finds a factor of 100 difference in amplitude and phase jumps of up to
$\pi$ radian between the lines which should originate in very similar regions of normal
stellar atmosphere. This unique behaviour is understood to be a result of vertical chemical
stratification (e.g., Kochukhov et al. 2006), combined with a rapid intrinsic height
variation of the magneto-acoustic pulsation waves propagating in stellar atmosphere.

The spatial filtering effect of chemical inhomogeneities opens interesting prospects for
horizontal and vertical resolution of pulsation modes in roAp stars, as can be done for no
other type of pulsating stars except for the Sun. The horizontal mapping of roAp pulsations
has already been performed with the help of an extended Doppler imaging technique (Kochukhov
2004). However, the vertical resolution of pulsation modes turns out to be considerably more
challenging because one has to abandon the standard single-layer approximation universally
adopted in detailed line profile modelling of non-radially pulsating stars (e.g.,
Briquet \& Aerts 2003; Schrijvers et al. 1997).

In an effort to get a better insight into the complex atmospheric dynamics of these stars, we
are conducting numerical, non-linear magneto-hydrodynamic (MHD) simulations of pulsational
wave propagation (Khomenko \& Kochukhov 2009). Here we use the resulting time-dependent
atmospheric structure and velocity field to predict line profile variations for roAp stars
accounting for vertical chemical stratification and using realistic line formation calculations.
This theoretical modelling represents a key step towards understanding the puzzling
complexity of the spectroscopic pulsations in roAp stars and eventually resolving 3-D
structure of their pulsation modes.

\begin{figure*}[!t]
\centering
\includegraphics[width=\textwidth]{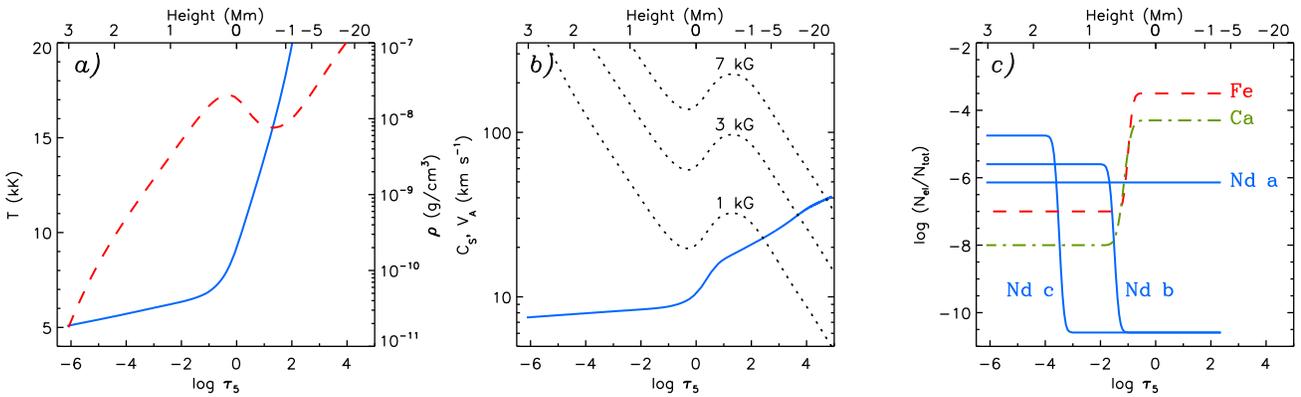}
\caption{
Initial model atmosphere and chemical stratification adopted for the MHD
calculations and line profile synthesis. {\bf a)} Temperature (solid line)
and density as a function of height and optical depth for an
unperturbed atmospheric model with $T_{\rm eff} = 7750$~K and $\log g = 4.0$.
{\bf b)} Depth-dependence of the sound (solid line) and Alfv\'en
(dotted lines) speeds for several values of the magnetic field strength.
{\bf c)} Vertical stratifications of Ca, Fe and three different
distributions of Nd employed in the spectrum synthesis.}
\label{atmos}
\end{figure*}

\begin{figure*}[!t]
\centering
\includegraphics[width=140mm]{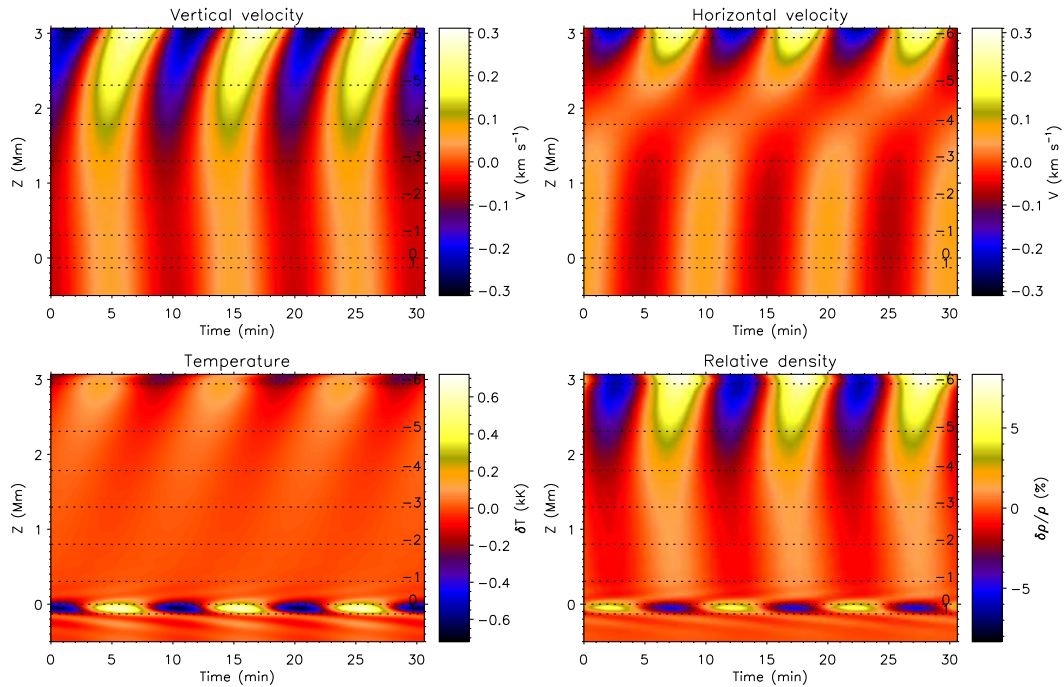}
\caption{
Height-time variation of the vertical and horizontal pulsation velocities
(top panels), temperature and density (bottom panels). These results
illustrate MHD simulations of the $P=10$~min pulsation at the intermediate
magnetic co-latitude of the $B_{\rm p}=3$~kG dipolar magnetic field.
The local field strength is $1.9$~kG and the field inclination is 44.8\degr\
with respect to the surface normal.
The horizontal dotted lines indicate the surfaces of constant optical depth
$-6\le\log\tau_{5000}\le1$.
}
\label{mhd}
\end{figure*}

\section{MHD simulations of roAp pulsations}

Our approach to modelling pulsational wave propagation in roAp
stars has been described in detail by Khomenko \& Kochukhov
(2009). Briefly, we perturb the lower boundary of a hydrostatic
LTE model atmosphere obtained with the LLmodels code (Shulyak et
al. 2004) and follow the resulting outward propagation  and
transformation of magneto-acoustic waves through the numerical
solution of the ideal MHD in two dimensions using the code
described in Khomenko \& Collados (2006) and Khomenko el al.
(2008). Adiabatic calculations are performed in  the
plane-parallel approximation for homogeneous magnetic field. A
series of such MHD runs for different magnetic field strengths and
inclinations represent a global structure of a low-degree roAp
pulsation mode.

In our new series of simulations presented here we have employed an initial static model atmosphere
with parameters $T_{\rm eff}=7750$~K and $\log g=4.0$, extended down to $\log \tau_{5000}\approx4$
to ensure the ratio of gas to magnetic pressure $\beta > 1$ at the lower boundary for the field
strengths of up to 7~kG. The temperature, density, sound and Alfv\'en speeds of our initial model
are illustrated in Fig.~\ref{atmos}. We consider pulsations with a period of 10~min and perform
calculations for 10 different co-latitudes of the $B_{\rm p}=3$~kG dipolar magnetic field.
Pulsational displacement at the lower boundary is assumed to have only a vertical component, which
is given by the $\ell=1$, $m=0$ spherical harmonic, aligned with the dipolar field axis.

Representative variations of the temperature, density and velocity
obtained in our MHD calculations are illustrated in Fig.~\ref{mhd}
for an intermediate magnetic co-latitude. One can see an important
role of the subphotospheric density inversion, the presence of
which leads to substantial variations of thermodynamic quantities.
At the same time, velocity amplitude rapidly increases as waves
propagate outwards and there is a gradual phase change as well.
The phase of pulsations at a given height also noticeably varies
across the stellar surface due to different superpositions of the
fast and slow magneto-acoustic wave components.

\begin{figure*}[!t]
\centering
\includegraphics[angle=90,width=137mm]{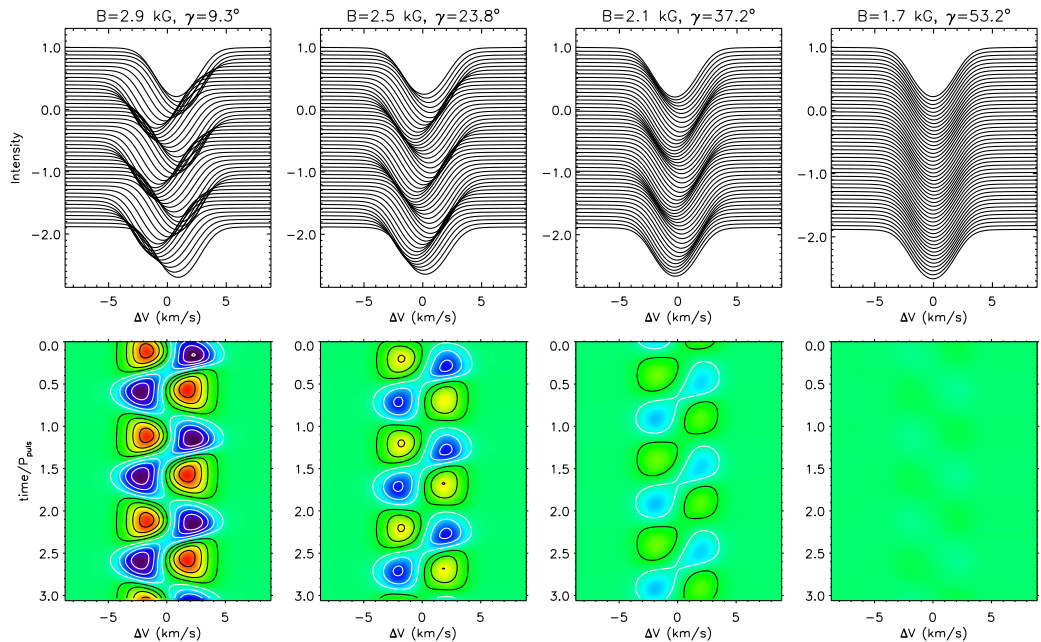}
\caption{Pulsational variation of the Nd~{\sc iii} 5851~\AA\ spectral
line for stratification ``Nd c'' at different magnetic co-latitudes,
characterized by a field strength $B$ and field inclination
$\gamma$. The upper panels show line profile variation for three pulsation cycles,
with spectral
lines corresponding to different pulsation phases offset vertically for
display purpose. The lower
panels illustrate the difference between time-resolved and average
spectra.
}
\label{lpv1}
\end{figure*}

\section{Line profile variations}

Simulations presented above and by Khomenko \& Kochukhov (2009) reveal pulsational wave behaviour
broadly consistent with the general depth-dependence of the radial velocity amplitude and phase
inferred in recent time-resolved spectroscopic studies of roAp stars. However, the ultimate quantitative
comparison with observations requires theoretical calculation of the line profile variation (LPV) using
the same spectral lines as studied by observers. Theoretical line profile modelling is also
essential for resolving the dispute about the nature of puzzling asymmetric profile variability
observed in some rare-earth lines (Kochukhov et al. 2007; Shibahashi et al. 2008).

Here we present the first preliminary results of the spectrum synthesis using our MHD models of pulsating
roAp atmospheres. We have chosen to model the Nd~{\sc iii} 5851~\AA\ spectral line as well as several
Fe~{\sc i}, {\sc ii} and Ca~{\sc i} lines. We adopt thermodynamic variables and velocity field from the
$\approx100$ snapshots of MHD simulations covering 3 pulsation cycles and calculate profile variations of
chosen spectral lines at different locations on the stellar surface. The spectrum synthesis calculations are
carried out with the help of a modified version of the SynthMag polarized radiative transfer code (Kochukhov 2007).

Line formation is treated accounting for typical vertical chemical inhomogeneities encountered in roAp stars
(e.g., Shulyak et al. 2009). These vertical chemical profiles are illustrated in Fig.~\ref{atmos}c. For Nd we explore three
distributions (a homogeneous one and two different distributions with a large overabundance in the upper layers) to mimic
diverse behaviour of different rare-earth ions. For Ca and Fe we adopt stratified distributions with both elements enhanced in
the lower atmospheric layers, as found in all cool Ap stars.

Fig.~\ref{lpv1} presents profile variations of the Nd~{\sc iii}
line at different magnetic co-latitudes. These profiles correspond
to the stratified distribution ``Nd c'' (see Fig.~\ref{atmos}c),
sampling the outermost atmospheric layers. Calculations show a
non-negligible effect of temperature and density variation, which
introduces a small asymmetry in the LPV pattern. At the same time,
magnetic field has a prominent influence on the pulsational
variability, concentrating it towards the magnetic pole.

\begin{figure*}[!t]
\centering
\includegraphics[width=137mm]{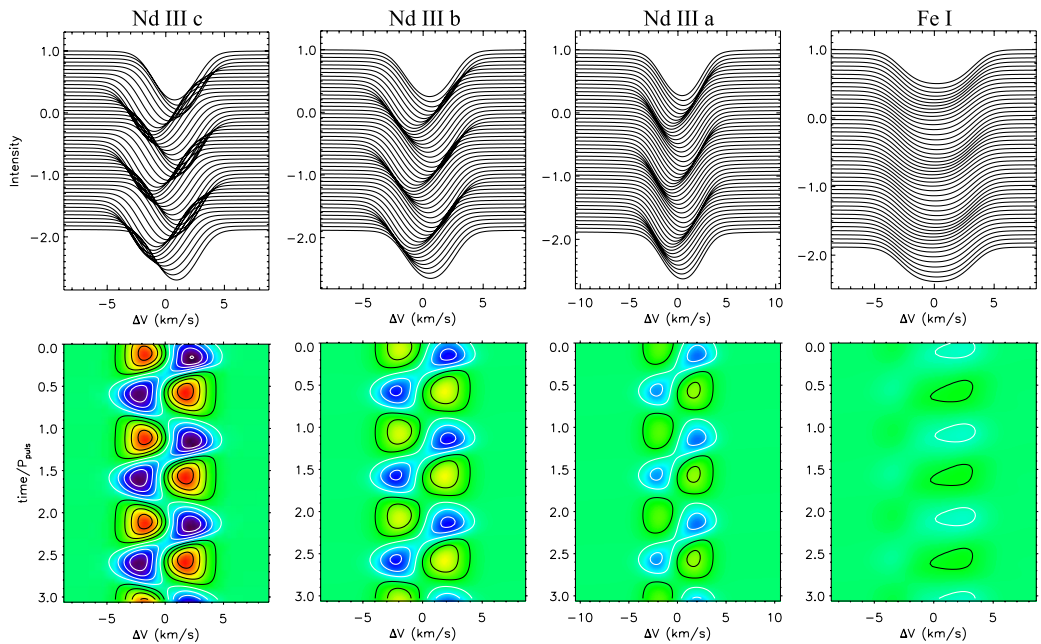}
\caption{
Comparison of the variation of Nd~{\sc iii} and Fe~{\sc i} spectral lines
at the location close to the magnetic/pulsational pole ($B=2.9$~kG,
$\gamma=9.3\degr$). The first three columns show profile variation of the Nd~{\sc iii}
line calculated for three different vertical distributions of Nd abundance
(see Fig.~\ref{atmos}). The last column shows behaviour of the Fe~{\sc i} line.
The format of this figure is similar to Fig.~\ref{lpv1}.
}
\label{lpv2}
\end{figure*}

Fig.~\ref{lpv2} compares theoretical LPV of the Nd line computed
with three different stratified abundance distributions of this
element with the variability predicted for a typical Fe~{\sc i}
line. As expected from the simulations results in Fig.~\ref{mhd},
we find a dramatic increase of the LPV amplitude towards the outer
atmospheric layers. Interestingly, our models predict that the
low-amplitude variability of the Fe and Ca lines formed in the
lower atmospheric layers is caused mainly by the strong
temperature and density fluctuations just below
$\log\tau_{5000}=0$. These results are verified by the spectrum
synthesis using constant thermodynamic structure and velocity
field from the MHD simulations. Since the variability of
temperature and density in the low atmospheric layers is related
to the subphotospheric density inversion, pulsational variations
of the Fe and Ca lines can be used to probe and explore the
structure of deep A-star atmospheric layers, hardly observable
otherwise.

\section{Conclusions}

\begin{itemize}
\item We have presented the first calculations of the pulsational LPV in roAp
stars based on detailed MHD models and realistic treatment of the
spectral line formation in magnetic, chemically stratified atmosphere.
\item In agreement with observations, our calculations show a large increase of the pulsational
amplitude and a gradual delay in phase from Fe and Ca to rare-earth spectral lines.
\item Even for the simple $\ell=1$, $m=0$ pulsation mode considered in our modelling the behaviour
of line profiles, as well as pulsational amplitude and phase, depends strongly on the local field
strength and inclination.
\item Pulsational changes of pressure and temperature contribute non-negligibly to the profile
variations of rare-earth lines and can dominate variations of Fe
and Ca lines.
\end{itemize}

\acknowledgements O.K. is a Royal Swedish Academy of Sciences
Research Fellow supported by grants from the Knut and Alice
Wallenberg Foundation and the Swedish Research Council. E.K. is
supported by the Spanish Ministerio de Ciencia e Innovaci{\'o}n
through projects AYA2007-63881 and AYA2007-66502.


\end{document}